\documentclass[english,5p]{elsarticle}
\usepackage[latin9]{inputenc}
\usepackage{babel}
\usepackage{float}
\usepackage{graphicx}
\usepackage{epstopdf}
\usepackage{amsmath}
\usepackage{esint}
\usepackage[unicode=true]
 {hyperref}
\begin{document}

\begin{frontmatter}{}

\title{Magneto-Optical Imaging of Vortex Domain Deformation in Pinning Sites}

\author{R. Badea, J. A. Frey, J. Berezovsky}

\address{Department of Physics, Case Western Reserve University, Cleveland,
Ohio 44106}
\begin{abstract}
We use a sensitive magneto-optical microscopy technique to image the magnetization response of micron-scale ferromagnetic disks to changes in applied magnetic field.  This differential technique relies on a modulated applied magnetic field which allows us to measure changes in magnetization $< 1\%$ with sub-micron resolution. The disks are magnetized in single vortex domains, with defects in the material serving to pin the vortex core at particular positions.  By applying a small AC magnetic field, we measure the deformation of the magnetization while the core remains pinned.  We can also characterize the strength of the pinning site by increasing the AC magnetic field to unpin the vortex core.  While pinned, we find that the magnetization away from the core reorients slightly to better align with an applied field.  Additionally, an applied field causes the pinned core itself to tilt in the direction of the field.  Once the field is large enough to unpin the core, this tilt disappears, and the core instead translates across the disk.
\end{abstract}

\end{frontmatter}{}

\section{Introduction}
The controlled movement of domain walls in micromagnetic structures is of increasing interest with promising applications as spintronic storage devices \cite{Parkin2008,Barnes2006} and logic elements\cite{Allwood2005}. As a result it is important to gain an understanding of domain wall dynamics. The presence of impurities and intrinsic material defects, found in all real samples, affects the mobility of domain walls causing them to be trapped at pinning sites\cite{Rahm2003,Uhlig2005,Compton2006,Min2010,VandeWiele2012}. Thin disks composed of soft magnetic materials have been shown to exhibit a ground-state vortex structure, which is characterized by a single large curl of in-plane magnetization and a central vortex core region of magnetization normal to the plane\cite{Cowburn1999,Shinjo2000,Wachowiak2002}. The high energy density of the central vortex region enhances local interaction with pinning sites \cite{Moura-Melo2008,Apolonio2009}, making these structures well suited for studying domain dynamics in the presence of disorder. 

Extensive work has been done on the interaction between magnetic vortices and artificially created pinning sites by means of micro-Hall magnetometry\cite{Rahm2003,Rahm2004,Rahm20042}, Lorentz transmission electron microscopy\cite{Uhlig2005}, and scanning x-ray transmission microscopy\cite{Kuepper2007}. Optical probing by means of time-resolved Kerr microscopy has allowed for the measurement of gyrotropic frequency suppression in the presence of naturally occurring pinning sites\cite{Compton2006,Compton2010,Chen2012}. Recent work has begun considering the magnetic field induced deformation of the vortex domain structure while it is pinned \cite{Burgess2014}, and indirect experimental evidence for the pinned vortex deformation has been seen\cite{Burgess2013}.

Here, we develop and use a sensitive magneto-optical technique to image the magnetization response to a change in magnetic field of both pinned and unpinned vortex domains, with sub-micron resolution.  By modulating both the probe laser and the applied magnetic field, we achieve sufficient sensitivity to map out the deformation of the vortex domain structure while it is trapped in a pinning site.  A similar technique was used in Ref.\cite{Chen2012} to measure temporal dynamics of magnetic domains -- here we focus on static magnetization maps.  We find that the in-plane magnetization away from the pinned vortex core deforms to better align with the applied field, as predicted by\cite{Burgess2014,Burgess2013}.  Additionally, we find that an asymmetry in the out-of-plane component of the magnetization arises when the vortex is deformed, which we attribute to a tilting of the pinned vortex core.  In both the pinned and unpinned cases, we also measure an out-of-plane magnetization component near the edge of the structures, likely due to the magnetostatic charge that accumulates on the adjacent sides of the magnetic disks.  These results provide a more detailed picture of the process of domain pinning and unpinning, and may lead to more accurate modeling of domain wall motion.

\section{Methods}

\begin{figure}[h]
\centering{}\includegraphics[scale=0.35]{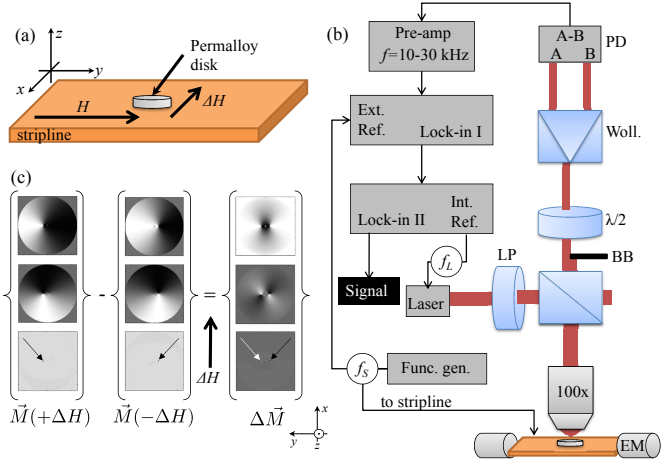}\protect\caption{{\footnotesize{}(a) An illustration of a Permalloy disk atop the gold microstripline. A constant field is produced by an electomagnet along the y-direction and an alternating field is produced along the x-direction by an alternating current through the stripline. (b) An illustration of the important optical and electronic components of the experiment. Here LP is a linear polarizer, BB is a beam blocker, $\lambda/2$ is a half-wave plate, Woll. is a Wollaston prism, PD is a balanced photo-detector, and EM is an electromagnet. (c) Simulated maps of the ${M}_{\alpha}$ component of magnetization with an applied field of $\pm\Delta H=600$~A/m, where $\alpha=x,y,z$ from top to bottom respectively. The difference of these maps is taken to generate the contrast map $\Delta M_\alpha$. Arrows indicate the small feature from the vortex core.}} 
\label{setup}
\end{figure}

A schematic of the sample is shown in Fig.~\ref{setup}(a).  A $100$-nm-thick gold stripline waveguide was patterned via photo-lithography and thermal evaporation on a sapphire substrate.  The stripline tapers to a $10~\mu$m width at the center.  Atop this central region of the stripline, $30$~nm thick Permalloy ($\mathrm{Ni_{0.81}Fe_{0.19}}$) disks (diameter $d = 1$ and $2~\mu$m) were fabricated via electron beam lithography, electron beam evaporation, and liftoff.  Naturally occurring defects in the Permalloy serve as pinning sites for the vortex core.

The experimental setup (Fig.~\ref{setup}b) is based on a standard scanning magneto-optical Kerr effect (MOKE) microscope~\cite{Freeman2002}.  A diode laser ($\lambda=660$~nm) is sinusoidally modulated at frequency $f_L=200$~Hz, and is linearly polarized and focused onto the sample through a 100x oil immersion objective (numerical aperture, $\mathrm{NA}=1.25$).  The sample is mounted on a three-axis piezo-driven stage, and is situated between the poles of an electromagnet which supplies a constant magnetic field $H_0$ in the y-direction. The reflected light is collected back from the objective, and the Kerr rotation is measured using a balanced photo-diode detector.  The measurement can be carried out by either collecting all of the reflected light, or by blocking the bottom, or left half of the reflected beam.  When all of the probe light is collected, the Kerr rotation measures the out-of-plane magnetization component $M_z$.  When half the beam is blocked, the collected light has reflected off the sample at an nonzero average angle of incidence, and therefore is also sensitive to the in-plane $M_x$ or $M_y$ magnetization component when blocking the bottom or left half, respectively.

An in-plane alternating magnetic field with amplitude $\Delta H$ was applied to the sample in the x-direction by running a square wave current through the waveguide at frequency $f_{S}=15$~kHz. The damping time for these magnetic domains are on the order of $t_d \sim 100$~ns \cite{Novosad2005}.  Because $f_S \ll t_d^{-1}$, we can assume that the magnetization alternates between two static configurations at $\pm \Delta H$.  There is a component of the MOKE signal arising from the difference in magnetization between $\pm \Delta H$ which will be modulated at frequency $f_S \pm f_L$.  The signal from the balanced photodiode detector was passed through an amplifier equipped with a band-pass filter from 10 kHz to 30kHz.  This filter removes any part of the signal not arising from the Kerr effect, such as reflection off of the waveguide when the laser spot is near the edge of the sample.  The signal is then sent to a lock-in amplifier with reference frequency $f_{S}$, and time constant $640~\mu$s. The output of the first lock-in amplifier is sent to a second lock-in amplifier with reference frequency $f_{L}$with time constant $200$~ms.

In a traditional scanning MOKE microscopy measurement only the probe laser is modulated, and after a single lock-in, magnetization over the surface of the sample $\vec{M}(\vec{\rho},\vec{H})$ is measured, where $\vec{\rho}$ is the two dimensional position vector \cite{Freeman2002}.  The different components of $\vec{M}$ are determined separately by blocking or unblocking the probe beam, as described above.  This technique is  particularly problematic near the edges of magnetic structures, or for small samples where the probe is always near an edge.  The presence of reflection from the underlying substrate and polarization effects due to the edges, in conjunction with sample vibration or position drift, translate into large sources of noise and signal drift.  These problems are largely eliminated with the addition of the alternating field $\Delta H$, and the second lock-in amplifier. In this case, we probe the change in the magnetization $\Delta\vec{M}(\vec{\rho},H,\Delta H)=\vec{M}(\vec{\rho},\vec{H}+\Delta\vec{H})-\vec{M}(\vec{\rho},\vec{H}-\Delta\vec{H})$.
The measured signal $\Delta \tilde{M}$ with the probe spot centered at position $\vec{\rho}$ is the convolution of the focused probe profile $I(\vec{\rho})$ and $\Delta M$:
\begin{multline}
\Delta\tilde{M}_{\alpha}(\vec{\rho},H,\Delta H)= \\
\int I_{\alpha}(\vec{\rho}-\vec{\rho}')\Delta M_{\alpha}(\vec{\rho'},H,\Delta H)d^{2}\rho',
\end{multline}
where $\alpha = x,y,z$.  In general, the focused probe profile depends on $\alpha$, because of the different beam blocking geometries.  $I_z$ is radially symmetric, whereas $I_x$ and $I_y$ are elongated in the x- and y- directions respectively, resulting in some distortion of the images.

A map of $\Delta \tilde{M}(\vec{\rho})$ is obtained by monitoring the signal while raster scanning the sample at a fixed $H$ and $\Delta H$.  The intensity of reflected light $R(\vec{\rho})$ is also monitored, and is used to normalize the signal to account for drift of laser intensity or differences in reflectivity.  Figure \ref{setup}c illustrates the interpretation of this measurement, for a simulated unpinned vortex.  At $+\Delta H$, the vortex core is shifted to the left, and at $-\Delta H$ the vortex is shifted to the right.  A map of $M_x$, $M_y$, and $M_z$ is shown for both cases.  The map of $\Delta \vec{M}$ represents the difference between these sets of images.  The actual measured images $\Delta \tilde{M}_\alpha$ are smeared out by the probe spot profile. 

In order to understand the measured maps of $\Delta \tilde{M}$, we have compared to the results of micromagnetic simulations (using the Object-Oriented Micromagnetic Framework, OOMMF \cite{OOMMF}).  The 30~nm thick, $2 ~\mu$m diameter magnetic disk is discretized with a $6\times6\times6$~nm grid ($333\times333\times5$ cells).  To model a pinning site, a 12-nm-radius region of the sample is set to a reduced saturation magnetization $M_s^{pin}= 0.85M_s$.  Initially in an idealized vortex, the magnetization is allowed to relax at different values of applied magnetic field, producing simulated maps of $\vec{M}(\vec{\rho},\vec{H})$.  We then calculate $\Delta \vec{M}$, and convolve with the diffraction-limited probe profile $I_\alpha$ to find $\Delta \tilde{M}_\alpha$.  Including the asymmetry of $I_{x,y}$ due to the partially blocked beam gives rise to some distortion of the simulated images, which can be seen in the experimental results.   For clarity, however, we will take $I_{x,y,z}$ to all be radially symmetric Gaussian spots in the simulated images presented below.

\section{Results and Discussion}

This technique provides both a straightforward means to characterize the pinning sites present in the sample, and to study how the magnetization changes with magnetic field both in the pinned and unpinned regimes.  First, we will characterize a pinning site, then discuss the behavior of the magnetization as it interacts with that site. 

\begin{figure}[h]
\centering{}\includegraphics[width=0.48\textwidth]{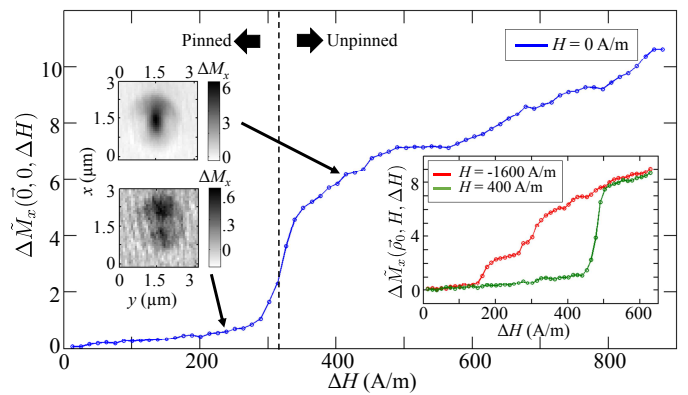}\protect\caption{{\footnotesize{}Magnetization response $\Delta \tilde{M}_{x}(\vec{0},0,\Delta H)$
at the center of a 2-$\mu$m-diameter and $30$-nm-thick Py disk. Also shown are magnetization response maps $\Delta\tilde{M}_{x}(\vec{\rho},0,\Delta H)$ taken at $\Delta H = 240$~A/m and $440$~A/m. Included in the inset are magnetization responses $\Delta\tilde{M}_{x}(\vec{\rho}_0,H,\Delta H)$ with a static offset of $H=-1600$~A/m and $400$~A/m, and $\vec{\rho}_0$ centered on the vortex core. }}
\label{barkhausen}
\end{figure}

When measuring $\Delta \tilde{M}$ at increasing values of $\Delta H$, a sudden jump in signal amplitude is observed as the vortex becomes unpinned, and the vortex core becomes free to move.  We can conveniently monitor this depinning transition by centering the probe spot at the vortex core position in the $\Delta \tilde M_x$ map, and sweeping $\Delta H$, as shown in Fig.~\ref{barkhausen}.  When the vortex is pinned, there is little change in the magnetization near the vortex core.  However, when $\Delta H$ is large enough to depin the vortex, the core can translate freely.  As the vortex core translates to both sides of its equilibrium position, $M_x$ changes from near $+M_s$ to $-M_s$ at this central point, resulting in a large signal in $\Delta \tilde{M}_x$.  

Figure~\ref{barkhausen} shows $\Delta\tilde{M}_x(\vec{0},0,\Delta H)$ from a $2~\mu$m diameter disk.  At $\Delta H < 320$~A/m, the signal at the center of the disk increases slowly due to the deformation of the vortex in the region overlapping with the probe spot, but away from the core.  Around $\Delta H = 320$~A/m, the signal increases rapidly as the vortex becomes unpinned.  At higher $\Delta H$, the curve shows (reproducible) behavior consisting of increasing signal and plateaux.  Here, the vortex is being shifted through the potential landscape away from the original pinning site, which includes other nearby pinning sites. The images in Fig.~\ref{barkhausen} show two scans of $\Delta\tilde{M}_x(\vec{\rho})$ at two values of $\Delta H$.  The image at $\Delta H =240$~A/m has lower signal than at $\Delta H = 440$~A/m, and less signal strength at the vortex core at the center of the disk.  

By changing the static field $H$, the vortex can be moved to different pinning sites, with different strength.  The inset to Fig.~\ref{barkhausen} shows scans of $\Delta \tilde{M}_x$ vs. $\Delta H$ at different values of $H$, centered on the vortex core equilibrium position $\vec{\rho}_0$.  In each data set we see a sudden increase in magnetization response indicating the vortex core has depinned from its position at $\Delta H=0$. The response taken at $H=-1600$~A/m demonstrates multiple distinct jumps indicating multiple depinning events with increasing field.

From the data in Fig.~\ref{barkhausen}, we can obtain an estimate of the energy $E_{pin}$ of the pinning potential.  To do so, we use a rigid vortex model (RVM), following Refs \cite{Burgess2014,Burgess2010,Guslienko2001}.  In the RVM, the vortex is specified by the distance $r$ from the center of the disk to the core. The total energy $E$ of the system, comprised of magnetostatic, exchange, and Zeeman energy, is expressed as a function of $r$.  In the absence of pinning, and with $r \ll R$ (the radius of the disk), the total energy can be expressed as 
\begin{equation}
\frac{E_0}{\mu_{0}M_{s}^{2}V}=\frac{\beta}{2}b^{2}-h\left(b-\frac{b}{8}^{3}\right)
\end{equation}
where $b=r/R$ the normalized radial distance, $M_{s}=8\times10^{5}$~A/m
is the saturation magnetization for Permalloy, $h=H/M_{s}$ is the
normalized applied field, $V$ is the volume of the entire structure,
and $\beta=0.0157$ is a parameter which we obtain by fitting the displacement of the core vs. $H$ in the micromagnetic simulations.  The minimum of $E_0$ vs. $h$ describes how the vortex moves freely in response to a magnetic field.  We then include a pinning site potential $E'$ in the total energy as $E = E_0 + E'$, with 
\begin{equation}
E'=-E_{pin} e^{-\frac{b^{2}}{2\gamma^{2}}}.
\end{equation}

Solution of this problem requires minimization of the total energy with respect to vortex core displacement. The width $\gamma$ and depth $E_{pin}$ of the pinning potential are adjusted such that depinning is observed at the field at which the depinning  step occurs in the data, resulting in a spectrum of solutions $E_{pin} (\gamma)$. $\gamma$ and $E_{pin}$ are further constrained by the slope of the pinned and unpinned signal, where increased curvature of the pinning potential results in a reduction of the pinned slope~\cite{Burgess2014, Burgess2013}. We find that the observed depinning transition at $H=318$~A/m corresponds to a depth of $E_{pin} \approx 1$~eV and width of $\gamma \approx 4$~nm. The other pinning sites shown in the inset to Fig.~\ref{barkhausen} are found to have depths ranging from $E_{pin} \approx 0.4 - 3.0$~eV and widths ranging from $\gamma  \approx 4 - 10$~nm.


\begin{figure}[h]
\centering{}\includegraphics[width=0.48\textwidth]{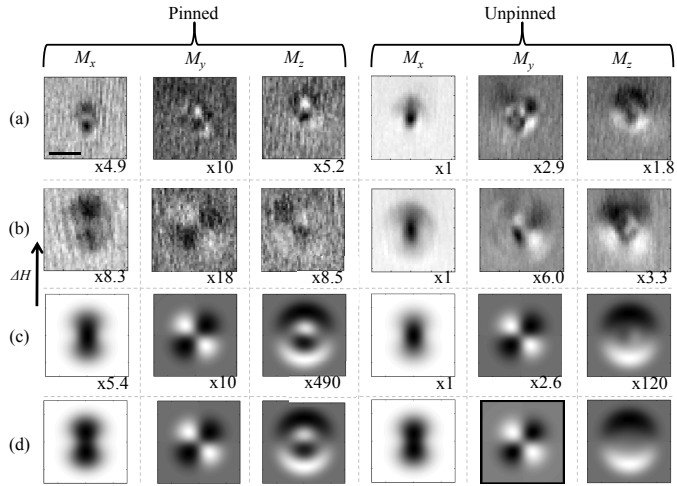}\protect\caption{{\footnotesize{}Magnetization response maps $\Delta\tilde{M}_{\alpha}(\vec{\rho},0,\Delta H)$) of (a) $1~\mu$m and (b) $2~\mu$m diameter $30$~nm thick Py disks in pinned and unpinned regimes along the three magnetization components ($\alpha=x,y,z$).The maps were generated using $\Delta H=240$~A/m (pinned) and $\Delta H=440$~A/m (unpinned) for the $2~\mu$m disk. Field strengths of $\Delta H=970$~A/m (pinned) and $\Delta H=1260$~A/m (unpinned) were used for the $1~\mu$m disk. All images are set to cover the same range of values after multiplication by the indicated scale factor.  (c) Simulated magnetization response maps generated using OOMMF for the $2~\mu$m disk with $\Delta H=200$~A/m (pinned) and $\Delta H=400$~A/m (unpinned). (d) Same as (c), but with the central 120-nm-radius region of simulated magnetization masked before convolution with the probe profile. }}
\label{images}
\end{figure}

Now we turn to the mapping of the magnetization response in the pinned and unpinned regimes.  Figures~\ref{images}(a) and (b) show magnetization response maps $\Delta\tilde{M}_{\alpha}(\vec{\rho},H=0,\Delta H)$ of $1~\mu$m and $2~\mu$m diameter $30$-nm-thick Py disks. The pinned scans were taken with $\Delta H$ below the jump seen in Fig.~\ref{barkhausen}  ($\Delta H=240$~A/m for the $2~\mu$m disks and $\Delta H=970$~A/m for the $1~\mu$m disk), and the unpinned scans were taken with $\Delta H$ above the jump ($\Delta H=440$~A/m for the $2~\mu$m disk and $\Delta H=1260$~A/m for the $1~\mu$m disk).  The $1~\mu$m and $2~\mu$m disks show qualitatively similar behavior in all cases, demonstrating the generality of these results.  In some cases, low signal-to-noise makes the pattern hard to discern, and the image may be more visible in one sample than the other. 

Figure~\ref{images}(c) shows the simulated magnetization response $\Delta \tilde{M}$ for the $2~\mu$m disk.  Here, a pinning site was included at the center of the disk, as described in the methods section.  In the simulation, motion of the vortex core is strongly suppressed for $\Delta H \leq 200$~A/m due to the presence of the pinning site.  When $\Delta H$ is increased to $400$~A/m, the vortex core is unpinned and exhibits a jump in position.
 
The relative magnitude of the images in Fig.~\ref{images}(a)-(c) are indicated by the scale factors below each image.  The gray scale is set to cover the same range of values for each image in (a), (b), and (c) by multiplying the data by the indicated factor.  In each case, the unpinned $\Delta M_x$ component has the largest signal range, and the others are scaled up to cover the same range.  The zero of the signal in each case can be seen as the background color.  In each case, we see the expected increase in signal (decrease in scale factor) going from the pinned to corresponding unpinned situation.  Additionally, we see that the $\Delta M_y$ component is smaller in magnitude than the $\Delta M_x$ component by a factor of $\sim 2-6$ in both measurements and in the simulation.  The simulated $\Delta M_z$ component is about 50 times smaller than the measured signal.  In the simulation, however, we do not take into account the differing MOKE efficiency for the $z$ component vs. the $x$ and $y$ components.  Given the relatively small angle of incidence of the incoming probe, we expect the experimental $z$ component to be significantly enhanced as compared to the $x$ and $y$ components.
  
The simulated images capture the essential features of the data, and allow us to interpret these images.  It should be noted that a portion of the $\Delta \tilde{M}_z$ component is also present in the $\Delta \tilde{M}_{x,y}$ images due to the probe angle of incidence $<~90^\circ$, which adds some distortion to the images.  The $\Delta \tilde{M}_{x,y}$ components also show distortion from the partially blocked probe beam.  For simplicity, neither effect is included in the simulated images shown.  We have also carried out the simulations with these distortions included, and better agreement with the experiment is obtained, at the expense of clarity.  
  
The primary difference in $\Delta \tilde{M}_x$ and $\Delta \tilde{M}_y$ between the pinned and unpinned cases is an increase in the magnitude of the signal as the vortex core becomes free to move.  Closer examination of the  $\Delta \tilde{M}_x$ images reveals that the unpinned image has increased weight near the center of the disk, as compared to the periphery.  In the pinned case, the signal arises from distortion of the vortex away from the core as it tries to align with the applied field.  Once the vortex core can move, the signal arises both from the distortion away from the core, and from the displacement of the core itself.  The contributions from the motion of the core and the distortion away from the core can be separated in the simulation by masking the vortex core in the simulation.  That is, we calculate $\Delta \tilde{M}_\alpha$ with the central $120$-nm-radius region of the $\Delta \vec{M}$ map set to zero, before the convolution with the probe spot.  Figure~\ref{images}(d) shows the masked simulated $\Delta \tilde{M}_\alpha$ images for the $2~\mu$m disk, in both the pinned and unpinned cases.  $\Delta \tilde{M}_y$ looks almost identical with the vortex core masked or unmasked, indicating that this component is due largely to deformation of the vortex away from the core.  The pinned $\Delta \tilde{M}_x$ images look similar with and without the mask, while the masked, unpinned $\Delta \tilde{M}_x$ does not show the increased signal at the center, attributed to motion of the vortex core.

The $\Delta \tilde{M}_z$ component shows the most striking difference between the pinned and unpinned regimes.  The pinned $\Delta \tilde{M}_z$ map shows an antisymmetric feature at the center of the disk aligned along $\Delta H$ (the vertical axis), whereas the unpinned $\Delta \tilde{M}_z$ map has a similar feature, but perpendicular to $\Delta H$ (the horizontal axis).    Additionally, both the pinned and unpinned case show an anti-symmetric feature at the top and bottom edges.  

\begin{figure}[h]
\centering{}\includegraphics[width=0.48\textwidth]{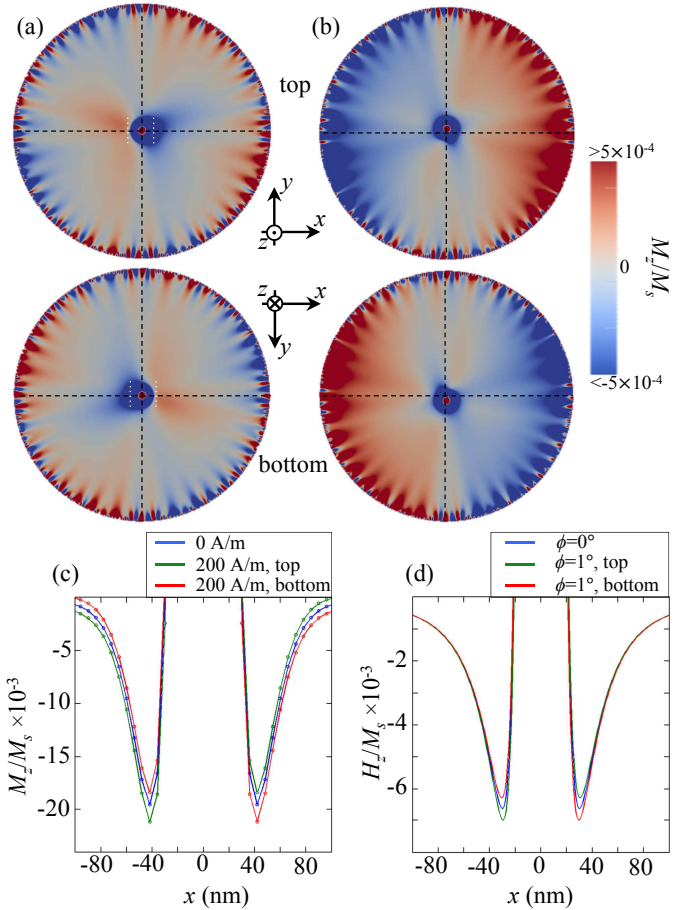}\protect\caption{{\footnotesize{}(a) and (b) Simulated $M_{z}$ component of magnetization of a $30$-nm-thick Py disk with $2\mu$m diameter and with a pinning site at the center of the disk, in a magnetic field $H_x=200$~A/m (a) and $H_x=400$~A/m (b).  (c) Line cut of $M_{z}/M_{s}$  taken at the center of the disk between the white dotted lines in (a). The line cuts were taken at $H=0$~A/m and $H_x=200$~A/m at the top and bottom surfaces of the disk. (d) The field $H_{z}/M_{s}$ produced by a dipole moment $|\vec{m}| = |\int \vec{M_{z}} d^{3}x|$ tilted from the $z$-axis by angle $\phi$ in the $x$-direction. The moment is positioned at the center of the disk and the field is calculated at the top and bottom surfaces. }}
\label{analysis}
\end{figure}

To understand the measured $\Delta \tilde{M}_z$ maps, in Fig.~\ref{analysis}(a) and (b) we plot the simulated $M_z$ on the top and bottom surfaces of the disk, with an applied magnetic field $H_x = 200$~A/m (pinned) and $H_x = 400$~A/m (unpinned).  Note that the color scale is set to a full range of $\pm 0.0005 M_s$ to show the relevant small effects.  In the vortex core $M_z \approx M_s$, so the color scale is completely saturated here, showing up as the central red dot.  The black dashed lines indicate the center of the disk.  The images of the top and bottom of the disk are flipped about the $x$-axis, as indicated by the coordinate axes.

The simulated images of $M_z$ show increasing out-of-plane magnetization at the edges of the disk with increasing applied field.  Comparing the sign of this effect on the top and bottom faces, this is a ``splaying'' of the magnetization with opposite sign on the opposing faces.  As the magnetic field is increased, the magnetization has a small component normal to the sides of the disk (not shown), as the magnetization begins to align with the field.  This normal component can be interpreted as magnetostatic ``surface charge'' accumulating on the sides of the disk.  The total energy is reduced when this surface charge spreads slightly onto the top and bottom faces resulting in outward pointing normal magnetization on one edge and inward pointing normal magnetization on the diametrically-opposite edge.  This explains the anti-symmetric feature occurring at the disk edges in the $\Delta \tilde{M}_z$ maps. The short length-scale modulations in $M_z$ seen near the edges of the simulated images arise from the discrete grid of the simulation.  A perfect circle would not show these modulations, though a similar effect may arise in practice from defects at the edge of the disk.  In any case, the length-scale of these modulations is too small to be seen in the experiments. 

The feature at the center of the the pinned $\Delta \tilde{M}_z$ map can be understood as a tilting of the pinned vortex core.  As seen in Fig.~\ref{analysis}(a), a ``halo'' , with $M_z \approx -0.02M_s$, surrounds the vortex core, with sign opposite to that in the core (appearing as a blue region).  This can be interpreted as the response of the magnetization to the magnetic field produced by the vortex core dipole moment. At $H=0$, the halo is symmetric about the vortex core (not shown).  As $H$ is increased to $200$~A/m, the halo becomes asymmetric along the applied field.  Moreover, the asymmetry is opposite on the top and bottom faces.  This asymmetry can be clearly seen in the line cuts of $M_z$ through the center of the disk, shown in Fig.~\ref{analysis}(c).  This can be explained by a tilting of the dipole moment of the vortex core, to partially align with the applied field.  For comparison, we can model the core as a magnetic dipole with moment $\vec{m}$, positioned at the center of the disk, halfway between the two faces.  Taking the magnetization distribution in the core as calculated in Ref.\cite{Hollinger2003}, we estimate $|\vec{m}|/M_s \approx 7900$~nm$^3$.  Figure~\ref{analysis}(d) shows the magnetic field produced by the dipole moment  at the top and bottom faces of the disk with $\vec{m}$ aligned along the z-axis, and tilted by $\phi=1^\circ$ with respect to the z-axis.  The asymmetry of $\sim 0.001 M_s$ produced by this tilt angle is consistent with the resulting deformation of the vortex magnetization seen in the simulation as magnetostatic charge is rearranged on the surfaces of the disk in response to this magnetic field.  This out-of-plane magnetization extending in the direction of $\Delta H$ gives rise to the observed feature near the center of the pinned $\Delta \tilde{M}_z$ map. Once the field has increased sufficiently to depin the vortex core, the energy is reduced by translating the vortex core instead of tilting it, as seen in Fig.~\ref{analysis}(b).  Here, the vortex core is no longer at the center of the disk, and the halo is largely symmetric once again.  This motion of the vortex core produces the feature at the center of the disk perpendicular to the $\Delta H$.

We can confirm the assignment of features in the $\Delta \tilde{M}_z$ maps by comparing the unmasked and masked images, in Fig.~\ref{images}(c) and \ref{images}(d).  In the pinned case, masking the vortex core from the simulated image does not greatly affect the image, indicating that the central feature is from the magnetization response away from the vortex core -- as explained above, it appears in the region surrounding the core in response to tilting of the core magnetization.  In contrast, the central feature in the unpinned $\Delta \tilde{M}_z$ map disappears completely when the vortex core is masked, consistent with the explanation that this feature arises from displacement of the vortex core.

\section{Conclusion}

We have employed a scanning magneto-optical microscopy technique to probe the response of vortex domains in micron-scale magnetic disks to changes in magnetic field.  By using a double-modulation scheme with two lock-in amplifiers, we are able to characterize the pinning sites present in the structure, and achieve measurements with sufficient sensitivity to observe changes in magnetization even while the vortex is ``pinned.''  We observe deformation of the vortex in the pinned state, as the magnetization away from the core is torqued by both the applied field, and the field from tilting of the pinned vortex core. These results yield a more detailed picture of how magnetic domains propagate in the presence of disorder.

\section*{References}
\bibliography{references}

\end{document}